\documentclass[aps,prl,twocolumn,nofootinbib,superscriptaddress,floatfix]{revtex4-2}

\usepackage[T1]{fontenc}
\usepackage[utf8]{inputenc}
\usepackage{amsmath,amssymb,mathtools,bm}
\usepackage[colorlinks=true,linkcolor=blue,citecolor=blue,urlcolor=blue]{hyperref}
\usepackage{microtype}
\usepackage{tikz}
\usepackage{float}
\usetikzlibrary{arrows.meta}

\newcommand{\mcA}{\mathcal A}

\newcommand{\bbI}{\mathbb I}
\newcommand{\OGr}{\mathrm{OGr}}
\newcommand{\GL}{\mathrm{GL}}
\newcommand{\Res}{\operatorname*{Res}}
\newcommand{\Vol}{\operatorname{Vol}}

\newcommand{\dlog}{\mathrm{d}\log}

\newcommand{\X}{\mathsf X}
\newcommand{\Sig}{\Sigma}
\newcommand{\Del}{\Delta}
\newcommand{\hS}{\widehat S}
\newcommand{\hT}{\widehat T}
\newcommand{\hU}{\widehat U}
\newcommand{\hc}{\widehat c}
\newcommand{\hSig}{\widehat\Sigma}
\newcommand{\hDel}{\widehat\Delta}

\begin{document}

\title{A Cosmological BCFW Bridge and Its Canonical Geometry}


\author{Aswini Bala}
\author{Sachin Jain}
\author{Vibhor Singh}

\affiliation{
Indian Institute of Science Education and Research, Pune, India\\
\textit{aswini.bala@students.iiserpune.ac.in},
\textit{sachin.jain@iiserpune.ac.in},
\textit{singh.vibhorrajesh@students.iiserpune.ac.in}
}

\date{\today}

\begin{abstract}
We build a BCFW-like recursion for cosmological correlators using the orthogonal Grassmannian. The key step is a bridge deformation that leaves all the Grassmannian constraints intact. The recursion relations are purely algebraic and avoid any spectral or radial integrals that usually appear in curved space. We obtain the gluon 4-point correlator using this bridge, producing two factorization poles, a total energy singularity coming from the three point building block, and a shifted energy singularity that emerges only after adding the two channels using recursion. We then show that the helicity stripped four-gluon correlator has the canonical form of a rectangle, where its boundaries represent various poles. We also describe the geometry of the gluon correlator in the Pfaffians as the natural coordinates in the Grassmannian. Using multiple deformation bridges, we also obtain the graviton four point correlator and observe a double-copy like relation at the level of the four point function, away from the physical cuts for each channel. We also extended the BCFW bridge formalism to the supersymmetric Grassmannian.
\end{abstract}

\maketitle

\paragraph{\textbf{Introduction:}}
One of the central outcomes of modern scattering theory is that tree-level amplitudes need not be computed  using Feynman diagrams. BCFW recursion \cite{Britto:2004ap, Britto:2005fq} can reconstruct them directly from lower-point on-shell data. Its power is that it is purely algebraic: the amplitude is simply a sum of residues at factorization poles, with no integrals to perform. Whether curved-space observables admit a similar algebraic recursion has till now remained far less clear.

Recursion relations in AdS and dS have so far been largely diagrammatic~\cite{Raju:2010by, raju2012new, PhysRevD.106.L121701}. 
The lower-point objects are sewn together using radial or spectral integrals over the 
exchanged state to obtain higher point function. This reflects the absence of a simple on-shell condition for 
internal lines. The result is recursive in spirit, but not as  clean as  flat space recursion. In this paper we show that, at least at four points, the 
cosmological correlator does admit a genuinely algebraic recursion. We work with the 
recently constructed three-dimensional cosmological Grassmannian~\cite{Arundine:2026fbr} (see 
also~\cite{De:2026shn,Bala:2026hdm,Bala:2026bdx,Huang:2026tsh,Arundine:2026myr}), where the correlator is an integral over a Grassmannian 
matrix subject to constraints. The deformation we introduce 
is an $O(n,n)$ bridge that acts simultaneously on the external spinor-helicity data and 
on the Grassmannian matrix. Crucially, it is an automorphism of the Grassmannian 
support: the constraints are preserved exactly. This implies the recursion can therefore be formulated directly on the support of the constraints. The four-point correlator is recovered as a sum of residues 
in the bridge parameter, the curved-space counterpart of the BCFW shift, rather 
than as an integral over an internal state.

One of the main focus of this paper is the study of  four gluon correlator in Grassmannian in $(- - + +)$ helicity which is given by \cite{Arundine:2026fbr}
\begin{equation}
\label{eq:mainA4}
\mcA_4^{\rm YM}(1^-2^-3^+4^+)
=
2\left({1\over \Sig}+{1\over \Del}\right){\X^2\over ST},
\qquad
\X=(12\bar3\bar4),
\end{equation}
where
\begin{equation}
\label{eq:sigdel}
\Sig=S+T+U,
\qquad
\Del=S+T-U.
\end{equation}
A central observation is that the bridge does not treat all poles democratically. 
Only $S=0$ and $T=0$ appear as finite poles in the shift parameter. The total-energy 
pole $\Sigma=0$ is inherited unchanged from the three-point building blocks, since 
$\Sigma$ is bridge-invariant. The shifted-energy pole $\Delta=0$ is not a pole in 
the shift at all; it emerges only after the two residues are combined, which is a unique feature of curved space.

These developments of BCFW and understanding of the poles naturally raises the question of geometry. In flat space, positive geometries,
associahedra, the amplituhedron, encode amplitudes as canonical forms. Every 
boundary is a factorization, collinear, or soft limit. For Cosmology the picture is different.  This is because the  total and shifted energy singularities are not surfaces where the correlator 
degenerates into lower-point objects. This raises an important  question:  is it possible to place singularities of genuinely different 
physical origin on the same footing?  We find that the stripped four-gluon 
correlator is a canonical form of a quadrilateral whose four edges are exactly the 
factorization boundaries $S=0$, $T=0$ and the cosmological energy boundaries 
$\Sigma=0$, $\Delta=0$. It is instructive to contrast this with flat space. In flat space at four points, the Mandelstam variables are not all independent, 
and the positive geometry is simply a line segment bounded by the two 
factorization poles $s=0$ and $t=0$. In contrast, in cosmology, the total energy is an independent kinematic 
variable, and this promotes the geometry to two dimensions. The result is 
a compact quadrilateral  whose facets are factorization and energy singularities on equal footing. The bounded region can also be understood in terms of Pfaffians, which are the natural coordinates in Grassmannian.

Finally, the bridge is not tied to gluons. It extends to scalars exchanging a gauge field and to the $\mathcal{N}=2$ super-Grassmannian, where the scalar correlator seeds 
the gluon correlator by supersymmetric projection. The scalar seed corresponds to a 
non-compact geometry; it is the helicity weight supplied by supersymmetry that closes 
the fourth edge and compactifies it. We also report a candidate four-graviton expression with the correct factorization and flat-space behavior  and showed that a double-copy-like relation exists away from the factorization limit at four points.

\paragraph{\textbf{Grassmannian support and the $O(n,n)$ bridge:}}
The bosonic cosmological Grassmannian data consist of a matrix $C=\{\bar1\bar2\cdots \bar n12\cdots n\} \in \OGr(n,2n)$ and spinor-helicity data
\begin{equation}
\Lambda=(\lambda_1,\ldots,\lambda_n,\bar\lambda_1,\ldots,\bar\lambda_n)^T,
\qquad
Q=\begin{pmatrix}0&\bbI_n\\ \bbI_n&0\end{pmatrix},
\end{equation}
which has to obey
\begin{equation}
\label{eq:support}
C\cdot Q\cdot C^T=0,
\qquad
C\cdot \Lambda=0.
\end{equation}
A schematic Grassmannian representation of the conformal correlator takes the form \cite{Arundine:2026fbr}:
\begin{equation}
\label{eq:grassrep}
\Psi_n(\Lambda)=\int {d^{n\times 2n}C\over \Vol(\mathbb{\GL}(n))}\,\delta(C\cdot Q\cdot C^T)\delta(C\cdot \Lambda)\,\mcA_n(C),
\end{equation}
where the rational function $\mcA_n(C)$ contains the dynamical information.

Let the deformation matrix\footnote{We choose to work with one real deformation parameter z, while multiple parameters are also allowed.} $B(z)\in O(n,n)$,such that
\begin{equation}
\label{eq:Onn}
B(z)^TQB(z)=Q.
\end{equation}
The bridge deformation is defined by
\begin{equation}
\label{eq:bridge-def}
\Lambda(z)=B(z)\Lambda,
\qquad
C(z)=CB(z)^{-1},
\end{equation}
such that,
\begin{align}
C(z)QC(z)^T
&=CB(z)^{-1}Q(B(z)^{-1})^TC^T=CQC^T,
\label{eq:preserve1}\\
C(z)\Lambda(z)&=CB(z)^{-1}B(z)\Lambda=C\Lambda.
\label{eq:preserve2}
\end{align}
implying that the deformation bridge preserves the defining constraints imposed on the Grassmannian. The recursion is therefore formulated directly on the cosmological Grassmannian support. Now, we show the effect of deformation on Grassmannian correlator, 
\begin{align}
    &\Psi_n(\Lambda(0)) \notag\\ &=\oint_{z=0} \frac{dz}{z} {\Psi}_n(\Lambda(z)) \notag \\ &= \int \frac{d^{n\times 2n} C}{\text{Vol}(\mathbb{GL}(n))} \delta(C\cdot Q\cdot C^T) \delta(C\cdot\Lambda) \oint \frac{dz}{z}\mathcal{A}(C(z)), \notag \\
         &=-\sum_{\substack{z=z_I}} \int dC~ \delta(C\cdot\Lambda)~ \frac{1}{{z}} Res_{z=z_I}(\mathcal{A}(C(z)) +B_{\infty}, 
\end{align}
where the sum $z=z_I$ is done over all the shifted factorized poles. 
\paragraph{\textbf{Four-point recursion:}}
 Let us now discuss the construction of a bridge for four points for which we choose the right-branch gauge of OGr(4,8),
\begin{equation}
\label{eq:right-branch}
C=(\bbI_{4\times4},-c_{4\times4}),
\qquad c_{ij}=-c_{ji}.
\end{equation}
We choose a bridge\footnote{The bridge $|ij\rangle$ refers to the spinors leg being deformed after the action of $B(z)$ on $\Lambda$.} $|1,3\rangle$, relevant for the $S$ and $T$ channel deformations in the current gauge as,
\begin{align}
\label{eq:B24}
&B_{24}(z)=
\begin{pmatrix}
A(z)&0\\0&(A(z)^{-1})^T
\end{pmatrix},
~~
A(z)=\bbI+z E_{24},\\
&\text{where},~~~~~~~~ E_{24}=
\begin{pmatrix}
0&0&1&0\\0&0&0&0\\0&0&0&0\\0&0&0&0
\end{pmatrix}.
\end{align}
Restoring the gauge gives
\footnote{Using
\[
B_{24}(z)^{-1}
=
\operatorname{diag}\!\big(A(z)^{-1},A(z)^T\big),
\]
one has
\[
C B_{24}(z)^{-1}
=
(I,-c)
\begin{pmatrix}
A(z)^{-1} & 0 \\
0 & A(z)^T
\end{pmatrix}
=
\big(A(z)^{-1},-cA(z)^T\big).
\]
This representative can be brought back to the right-branch gauge by the allowed equivalence C$\sim$ g$\cdot$C| $g=A(z)\in GL(4)$ :
\[
\big(A(z)^{-1},-cA(z)^T\big)
\overset{\mathrm{GL}(4)}{\sim}
\big(I,-A(z)cA(z)^T\big).
\]
Hence \(\hat c(z)=A(z)cA(z)^T\).}
\begin{equation}
\label{eq:cz}
\hat c(z)=A(z)cA(z)^T,
\end{equation}
so that
\begin{equation}
\label{eq:cshifts}
\hc_{12}=c_{12}-zc_{23},
\qquad
\hc_{14}=c_{14}+zc_{34},
\end{equation}
with all other $c_{ij}$ remaining unchanged.

\noindent The four-point Grassmannian Mandelstam are
\begin{align}
S&=(\bar1\bar212)=c_{13}c_{24}-c_{14}c_{23},
\label{eq:Sdef}\\
T&=(\bar1\bar414)=c_{13}c_{24}-c_{12}c_{34},
\label{eq:Tdef}\\
U&=(\bar1\bar313)=-c_{14}c_{23}-c_{12}c_{34}.
\label{eq:Udef}
\end{align}
Unlike flat-space Mandelstam's, they do not obey $S+T+U=0$. Under this bridge deformation, eq \eqref{eq:cshifts}, we get
\begin{equation}
\label{eq:STUshift}
\hS(z)=S-zq,
~~
\hT(z)=T+zq,
~~
\hU(z)=U,
\end{equation}
where $q=c_{23}c_{34}$.
We also get
\begin{equation}
\label{eq:sigma-invariant}
\begin{aligned}
\hSig(z) &= \hS(z)+\hT(z)+\hU(z)=\Sig,\\
\hDel(z) &= \hS(z)+\hT(z)-\hU(z)=\Del.
\end{aligned}
\end{equation}
Under this deformation, the factorization poles are situated at,
\begin{equation}
\label{eq:bridge-poles}
\hS(z_S)=0
\implies z_S = \frac{S}{q}, \quad
\hT(z_T)=0 \implies z_T={-T\over q}.
\end{equation}
At these poles,
\begin{align}
&z=z_S:
&&\hS=0,
&&\hT=S+T,
&&\hU=U,
\label{eq:zS-values}\\
&z=z_T:
&&\hT=0,
&&\hS=S+T,
&&\hU=U.
\label{eq:zT-values}
\end{align}

\paragraph{\textbf{Gluon four point with gluon exchange}:}
The recursion takes the usual BCFW form: the four-point object is reconstructed from products of two three-point objects, summed over the helicity of the internal line. In the present case,
\begin{equation}
\label{eq:bridge-recursion-general}
A_{4}^{- - + +}(C)
=
{N_S^{\rm shift}\over S}
+
{N_T^{\rm shift}\over T}
+
B_\infty ,
\end{equation}
where
\begin{equation}
\label{eq:NS-helicity-sum}
N_S^{\rm shift}
=
\left[
\sum_{h=\pm}
A_3^{\rm YM}(\hat 1^-,\hat 2^-,\hat I^h)\,
A_3^{\rm YM}(\hat 3^+,\hat 4^+,(-\hat I)^{-h})
\right]_{\hat S(z)=0}
\end{equation}
and similarly for $N_T^{\rm shift}$.
Here, hats denote the bridge-deformed kinematics, and the internal leg is glued with opposite helicities on the two three-point factors. The condition \(\hat S(z)=0\) fixes the bridge parameter to \(z=z_S\), so the gluing is algebraic in the Grassmannian rather than an integral over an internal radial or spectral variable. We should also emphasize that we have taken $B_{\infty}=0$ following the Witten-diagram analysis \cite{raju2012new}. This is analogous to flat-space Yang-Mills, where $B_\infty = 0$  is established by Feynman-diagram power counting and then imported into the Grassmannian framework rather than derived within it. An intuitive way to understand this is that the large-$z$ regime, which is equivalent to the high energy regime, locally makes the AdS/dS behave like flat space. So the Yang-Mills cubic and quartic vertices carry the same helicity-dependent $z$-scaling as in flat space, giving $B_{\infty}=0$ for $({-}\,{-}\,{+}\,{+})$ configuration irrespective of the sign of the cosmological constant. In fact, a better understanding of the boundary terms in the Grassmannian is required.

Before evaluating the recursive numerator, it is important to illustrate how the two three-point factors are uplifted as functions on the four-point Grassmannian. See Appendix B of \cite{Arundine:2026fbr} for a similar discussion.
For example, for the s-channel, the uplift means that the product
\[
\mathrm{OGr}(3,6)_L \times \mathrm{OGr}(3,6)_R
\]
is first embedded into \(\mathrm{OGr}(4,8)\) at the support of S=0. The internal line $\lambda_s,\bar\lambda_s$ is glued once, and its redundant \(\mathrm{GL}(1)\) scaling is divided out. After this embedding, the two three-point Grassmannian give the \(S\)-channel contribution to the four-point object. 

\noindent The uplifted product using the three point gluon correlator with leading interaction given in eq (3.7) of \cite{Arundine:2026fbr}, the shifted factorized expression gives
\begin{equation}
\label{eq:NS-hatted}
N_S^{\rm shift}
=
\left[
{4\X^2\over
(\hT+\hU)(\hT-\hU)}
\right]_{z=z_S}.
\end{equation}

\noindent Using eq \eqref{eq:zS-values}, we obtain
\begin{equation}
\label{eq:NS-final}
N_S^{\rm shift}={4\X^2\over \Sig\Del},
\end{equation}
where the numerator $\X=(12\bar3\bar4)$ is unchanged by this bridge in the chosen gauge.  Similarly, for the t-channel
\begin{equation}
\label{eq:NT-hatted}
N_T^{\rm shift}
=\left[{4\X^2\over (\hS+\hU)(\hS-
\hU)}\right]_{z=z_T}
={4\X^2\over \Sig\Del}.
\end{equation}
Substituting into eq \eqref{eq:bridge-recursion-general},
\begin{align}
A_{4}^{- - + +}
&={4\X^2\over \Sig\Del}\left({1\over S}+{1\over T}\right)
={4\X^2(S+T)\over ST\Sig\Del}.
\end{align}
Since
\begin{equation}
\label{eq:SplusT}
S+T={\Sig+\Del\over 2},
\end{equation}
we obtain the required correlator,
\begin{equation}
\label{eq:rec-final}
\boxed{A_{4}^{- - + +}
=2\left({1\over \Sig}+{1\over \Del}\right){\X^2\over ST}.}
\end{equation}
Note that the above correlator obtained by bridging, gives either 1-3 leg or 2-4 leg discontinuity, depending on the pole enclosed in the  $\tau$ plane, while converting to momentum space \cite{Arundine:2026fbr}.

Several features are worth emphasizing. First, only $S=0$ and $T=0$ are bridge poles which appears at the finite value of parameter $z$. Second, $\Sigma$ is invariant under the bridge, and it comes from the three-point Grassmannian correlators, not from a finite pole in the bridge parameter ($z$) plane. Third, $\Del$ is also invariant under this bridge but appears in the final answer only after evaluating the shifted three-point products on the factorization poles and combining the two channels. Thus, $\Del$ and $\Sigma$ are the physical singularities of the correlator, not an independent BCFW pole in $z$. 

Note that we can also obtain another form of the gluon correlator \cite{Arundine:2026fbr}, which satisfies the Kleiss-Kuijf relation by performing multiple deformations table \ref{tab:Table1} and adding their contributions in the S and T channels.
\begin{table}[H]
\centering
\begin{tabular}{|@{\hspace{0.3cm}}c@{\hspace{0.3cm}}|@{\hspace{0.3cm}}c@{\hspace{0.3cm}}|@{\hspace{0.3cm}}c@{\hspace{0.3cm}}|@{\hspace{0.3cm}}c@{\hspace{0.3cm}}|}
\hline
Bridge & $\hat{S}$ & $\hat{T}$ & $\hat{U}$ \\
\hline
$|1,3\rangle$ & $S-zq_{13}$ & $T+zq_{13}$ & $U$ \\
$|1,4\rangle$ & $S-zq_{14}$ & $T$ & $U+zq_{14}$ \\
$|1,2\rangle$ & $S$ & $T-zq_{12}$ & $U+zq_{12}$ \\
\hline
\end{tabular}
\caption{Multiple Deformation Bridges}
\label{tab:Table1}
\end{table}

\noindent which leads to:
\begin{align}
   A_{4}^{- - + +} ==\frac{(12\bar3\bar4)^2}{\Sigma}
\left[\frac1{S\Delta_U}+\frac1{T\Delta_U}-\frac1{S\Delta_T}-\frac1{T\Delta_S}\right],\notag
\end{align}
\begin{align}
   \boxed{A_{4}^{- - + +}= \frac{(12\,\bar3\,\bar4)^2}{2 ST} \left[ \frac3{\Sigma}+\frac1{\Delta_U}-\frac1{\Delta_S}-\frac1{\Delta_T}\right].}
\end{align}
where $\Delta_U=S+T-U, \Delta_T=S+U-T ~\text{and}~ \Delta_S=T+U-S.$
 This form of gluon correlator, when converted to momentum space, gives 1,2,3 leg discontinuity \cite{Arundine:2026fbr}.

\paragraph{\textbf{Positive Geometry and Canonical Form:}}
In the positive-geometry perspective \cite{Arkani-Hamed:2012zlh,Arkani-Hamed:2017mur,Arkani-Hamed:2024jbp,Arkani-Hamed:2017fdk}, the singularities of an amplitude are encoded as boundaries of a region in kinematic space. The canonical form \cite{Huang:2013owa,Arkani-Hamed:2012zlh} associated with this geometry is defined such that it has poles on these boundaries,  and its residues reproduce the simpler objects associated with each boundary. Unlike the flat space case, for cosmological correlators, the relevant boundaries are not just ordinary factorization channels, but also total-energy and shifted-energy singularities. This makes the canonical-form description interesting: it organizes both types of singularities within one geometric structure.   For the present four-point correlator, the natural kinematic coordinates are
\begin{equation}
\label{eq:xy}
x={S\over S+T},
\qquad
 y={U\over S+T}.
\end{equation}
The singularities are located in the $(x,y)$ plane 
\begin{equation}
\label{eq:pole-map}
\begin{aligned}
S=0 &\leftrightarrow x=0,
\qquad
T=0 \leftrightarrow x=1,\\
\Sig=0 &\leftrightarrow y=-1,
\qquad
\Del=0 \leftrightarrow y=+1.
\end{aligned}
\end{equation}

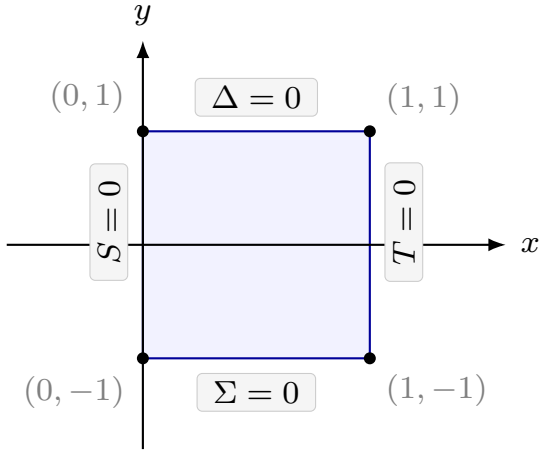
\begin{tikzpicture}[
    scale=1.5,               
    transform shape,         
    >=Latex,
    every node/.style={font=\footnotesize},
    labelbox/.style={
        draw=gray!40,
        fill=gray!8,
        rounded corners=2pt,
        inner xsep=4pt,
        inner ysep=2pt,
        font=\footnotesize
    },
    leader/.style={
        gray!60,
        dash pattern=on 2pt off 2pt,
        line width=0.4pt
    }
]

\coordinate (A) at (0,1);   
\coordinate (B) at (2,1);   
\coordinate (C) at (2,-1);  
\coordinate (D) at (0,-1);  

\fill[blue!5] (A) -- (B) -- (C) -- (D) -- cycle;
\draw[blue!60!black, thick] (A) -- (B) -- (C) -- (D) -- cycle;

\draw[->, thick] (-1.2,0) -- (3.2,0) node[right] {$x$};
\draw[->, thick] (0,-1.8) -- (0,1.8) node[above] {$y$};


\foreach \p in {A,B,C,D} \fill (\p) circle (1.5pt);
\node[above left=1pt, gray] at (A) {\footnotesize $(0,1)$};
\node[above right=0.5pt,  gray] at (B) {\footnotesize $(1,1)$};
\node[below right=0.5pt,  gray] at (C) {\footnotesize $(1,-1)$};
\node[below left=1pt, gray] at (D) {\footnotesize $(0,-1)$};


\node[labelbox] at (1,1.3) {$\Delta=0$};

\node[labelbox, rotate=90] at (2.3,0.2) {$T=0$};

\node[labelbox] at (1,-1.3) {$\Sigma=0$};

\node[labelbox, rotate=90] at (-0.3,0.2) {$S=0$};

\end{tikzpicture}

Thus, these four singular loci bound a natural rectangle\footnote{ In (S, T, U) space the geometry is a quadrilateral.} in the \((x,y)\) plane, as shown in the above figure,
\[
R=\{(x,y):0<x<1,\,-1<y<1\}.
\]
Its canonical form is\footnote{With the homogeneous lift fixed by the degree \(-3\) energy scaling of the
stripped correlator, the canonical function of the rectangle gives
\begin{equation}
\label{eq:canonical-function-lift}
\omega_R(x,y)
={2\over x(1-x)(1-y^2)},
~
{\mcA_4^{\rm YM}\over \X^2}
={2\over (S+T)^3}\,
\omega_R\!\left({x},{y}\right).
\end{equation}
Substituting the definitions of \(x\) and \(y\), this becomes
\begin{equation}
\label{eq:canonical-function-amplitude}
{\mcA_4^{\rm YM}\over \X^2}
=2\left({1\over \Sig}+{1\over \Del}\right){1\over ST}.
\end{equation}}
\begin{equation}
\label{eq:rect-form}
\Omega_R
=\dlog {x\over 1-x}\wedge \dlog {1+y\over 1-y}
={2\,dx\wedge dy\over x(1-x)(1-y^2)}.
\end{equation}
The residues satisfy
\begin{align}
\Res_{x=0}\Omega_R&={2dy\over 1-y^2},
&
\Res_{x=1}\Omega_R&=-{2dy\over 1-y^2},
\label{eq:xresforms}\\
\Res_{y=1}\Omega_R&=-{dx\over x(1-x)},
&
\Res_{y=-1}\Omega_R&={dx\over x(1-x)}.
\label{eq:yresforms}
\end{align}
\noindent
We choose the orientation \(dx\wedge dy\). With this choice, the residues in
eq \eqref{eq:xresforms} and eq \eqref{eq:yresforms} are canonical forms of the four edges of the rectangle.
Taking one more residue at any corner gives \(\pm 1\). The signs depend on
the orientation, while the unit magnitude fixes the overall normalization of
\(\Omega_{\mathcal R}\). The positive domain is
\begin{equation}
0<x<1,\qquad -1<y<1 .
\end{equation}
Equivalently, after fixing \(S+T>0\), this domain can be written as
\begin{equation}
S>0,\qquad T>0,\qquad \Sigma>0,\qquad \Delta>0 .
\end{equation}
 
\textbf{\small{Pfaffian coordinates:}} Pfaffian coordinates are the natural coordinates on the Grassmannian. So, one may also use the Pfaffian coordinates to give a geometrical interpretation of Gluon correlator. 

The Pfaffians of the sub-matrix eq \eqref{eq:right-branch} C:
   \begin{align}
       Pf(c_{4\times4})= \{p_{\phi}, p_{ij}, p_{1234}\},
   \end{align}
where $p_{\phi}=1$, $p_{ij}=c_{ij}$ and $p_{1234}= p_{12} p_{34} - p_{13} p_{24}+p_{14} p_{23}$. Let,
\begin{equation}
\label{eq:ABD}
A=c_{13}c_{24}= p_{13}p_{24},\,
B=c_{14}c_{23}= p_{14} p_{23},\,
D=c_{12}c_{34}= p_{12} p_{34}.
\end{equation}
Then
\begin{align}
\label{eq:ABD-poles}
S&=A-B=p_{13}p_{24}-p_{14}p_{23},\notag \\
T&=A-D=p_{13}p_{24}-p_{12}p_{34},\notag \\
\Sig&=2(A-B-D)= -2 p_{1234} ,\notag \\
\Del&=2 p_{13} p_{24}.
\end{align}
The helicity stripped correlator becomes
\begin{align}
\label{eq:quadform}
{A_4^{\rm YM}\over \X^2}
&={2A-B-D\over A(A-B)(A-D)(A-B-D)}, \notag \\
&= \frac{p_{13}p_{24}-p_{1234}}{-2p_{13}p_{24}(p_{13}p_{24}-p_{14}p_{23})(p_{13}p_{24}-p_{12}p_{34})p_{1234}}
\end{align}
This is the {projective canonical function
of the quadrilateral in the Grassmannian space, with the poles giving the boundaries.



 Thus the four-point gluon correlator is described by a single positive domain whose facets have different physical origins. This differs from flat-space four-point kinematics, where the
corresponding geometry is only a line segment bounded by the two factorization channels.

\paragraph{\textbf{Why isolated exchange geometries are not enough:}}
The above construction  clarifies that a single isolated exchange channel does not  define a compact positive geometry.  For example, keeping only an $S$-channel exchange produces only two singular hyperplanes, $S=0$ and $\Sig=0$, leaving the two-dimensional kinematic region non-compact.  The four-gluon correlator is a clean first example because all its relevant singularities are simple. This allows the factorization poles and the cosmological energy poles to be described by a single ordinary canonical form.

\paragraph{\textbf{Colored Scalar correlator with gauge-field exchange:}}
The bridge $|1,3\rangle$ applies equally to scalars exchanging a gauge field.
The shifted three point products evaluates to
$N^{\rm shift}_{S}=(\hat T - U)/(\hat T + U)\big|_{\hat S=0}
 = \Delta/\Sigma$,
and identically for the $T$-channel, giving
\begin{equation}\label{scalA4}
  {A}_{4}
  = \frac{\Delta}{\Sigma}\!\left(\frac{1}{S}+\frac{1}{T}\right).
\end{equation}
This simple example already demonstrates that the total-energy structure emerges dynamically from
the shifted three-point data.
\paragraph{\textbf{Scalar with graviton exchange:}} For scalar correlators involving graviton exchange, the bridge fixes the exchange singularities up to boundary/contact terms. It is known that due to the nature of the scalar-graviton interaction, there will be a non-trivial contribution from the boundary term.  

Using the deformation bridge $|2,4\rangle$, one can obtain the s-channel as,
\begin{align}\label{eq:Owithgexchangebridge}
        \langle OOOO\rangle_{s,\text{bridge}} = \frac{1}{S}\frac{(T-U)^2}{\Sigma^2}+\frac{S}{\Sigma^2}+ 2 \frac{(T-U)}{\Sigma^2}.
    \end{align}
The contributions from the remaining channels are obtained using the corresponding deformations, and their sum yields the complete four-point function:
 \begin{align}\label{oooo}
     &\langle OOOO\rangle_{\text{bridge}}\notag\\ &= \frac{1}{S}\frac{(T-U)^2}{\Sigma^2}+\frac{1}{T}\frac{(S-U)^2}{\Sigma^2}+\frac{1}{U}\frac{(S-T)^2}{\Sigma^2}+\frac{1}{\Sigma},
\end{align}
where the third term in eq \eqref{eq:Owithgexchangebridge} cancels after adding all the channel. It is easy to show that \eqref{oooo} has the correct flat space limit. However, as expected the \eqref{oooo} differs from one presented in \cite{Arundine:2026fbr}
\begin{align}
\langle & OOOO\rangle_{P_2}
\notag \\ &=
{C}\left[
{ {2S\over 3}\,
P_2\!\left({T-U\over S}\right)
\over
\Sigma^2}
+
{ {2T\over 3}\,
P_2\!\left({U-S\over T}\right)
\over
\Sigma^2}
+
{ {2U\over 3}\,
P_2\!\left({S-T\over U}\right)
\over
\Sigma^2}
\right],
\label{eq:OOOOP2}
\end{align}
by a 
contact term 
\begin{align}
    -\frac{2}{3 \Sigma}, 
\end{align}
which is not seen by the BCFW deformation. 
\paragraph{\textbf{Four graviton with graviton exchange:}}For pure spin two, applying similar bridge logic, \footnote{One should note that the current deformation bridge $|2,4\rangle$, which is discussed in the main text, only deforms the S and T minors. In order to capture the U pole of the graviton 4-point correlator, we have to do multiple deformations from Table (\ref{tab:Table1}) and add the terms together to get all the exchanges.}, the four-graviton helicity configuration gives all the exchange candidates.
\begin{align}\label{eq:four-graviton-discussion}
A_{4,\rm grav}^{--++}=
{X^4\over \Sigma^2}
\left[
{1\over S}
\left(
{1\over \Delta_U^2}
+
{1\over \Delta_T^2}
\right)
+
{1\over T}
\left(
{1\over \Delta_S^2}
+
{1\over \Delta_U^2}
\right)
\right. \notag \\
\hspace{4.2cm}\left.
+
{1\over U}
\left(
{1\over \Delta_T^2}
+
{1\over \Delta_S^2}
\right)
\right].
\end{align}
This expression has the correct spin-two factorization residues and the expected flat-space limit. One should also emphasize that the candidate answer may not be complete and may require a boundary term. In principle, one can perform spin-2 Legendre completion as done for the scalar four point with graviton exchange eq \eqref{eq:OOOOP2}.  A direct comparison with the full Euclidean spinor-helicity answer should fix these issues; However, this matching is more subtle. One possible route would be to use SUSY to relate the four graviton correlator to a simpler correlator for which spinor helicity matching would be easier.  It is also interesting to note that, due to the presence of the double pole structure, there is no obvious way to write a canonical form.

\textbf{\small{Double Copy:}} One can observe that, at the level of the three point case, the graviton and gluon have a trivial double copy relation. With that, one can ask if the same exists for the four point case as well. We found a hint towards the presence of a double copy like relation at the level of the deformed four point Grassmannian correlator, away from any of the physical cuts, for each channel at each bridge deformation \footnote{A double copy relation was also found at four points in Twistor space at the special kinematics regime \cite{Ansari:2025fvi} which led to at the cut S=0.}.  
For example:  With $|13\rangle$ bridge deformation, at S and T channel
\begin{align}
    N^{shift|13\rangle}_{S, J=2} = \frac{X^4}{\hat\Sigma^2\hat\Delta_U^2}\bigg|_{z=z_s} = (N^{shift|13\rangle}_{S, J=1})^2, \notag \\
    N^{shift|13\rangle}_{T, J=2} = \frac{X^4}{\hat\Sigma^2\hat\Delta_U^2}\bigg|_{z=z_T} = (N^{shift|13\rangle}_{T, J=1})^2.
\end{align}
Similarly, with $|14\rangle$ and $|12\rangle$ bridge, deformed S and T channel respectively gives, 
 \begin{align}
    N^{shift|14\rangle}_{S, J=2} = \frac{X^4}{\hat\Sigma^2\hat\Delta_T^2}\bigg|_{z=z_s} = (N^{shift|14\rangle}_{S, J=1})^2, \notag \\
    N^{shift|12\rangle}_{T, J=2} = \frac{X^4}{\hat\Sigma^2\hat\Delta_S^2}\bigg|_{z=z_T} = (N^{shift}_{T, J=1})^2.
\end{align}
An exact double copy relation or color kinematics duality would be very interesting result \cite{XYZ:2026new4}. 

\paragraph{\textbf{Bridge deformation in SUSY:}}
We can extend the bridge deformation to supersymmetric Grassmannian~\cite{Bala:2026bdx},
whose wavefunction representation is
\begin{align}
  \mathbf\Psi_n
  = \!\int\!\frac{d^{n\times 2n}C}{\mathrm{Vol}\,GL(n)}\;
\delta(CQC^T)\,\delta(C\Lambda)\, \hat\delta(C\,\Xi)\, \mathcal{F}\!(C),
\end{align}
where $\hat\delta(C\cdot\Xi)$ encodes superconformal
and $R$-symmetry constraints through an operator-valued
fermionic delta function.
Along with non-supersymmteric deformation, we also deform $\Xi\to B(z)\Xi$ in the same $O(n,n)$
representation as~$\Lambda$, which implies  the bridge preserves all three supports:
$$C(z)QC(z)^T\!=CQC^T,\;C(z)\Lambda(z)\!=C\Lambda,\;
C(z)\Xi(z)\!=C\Xi.$$
Using this supersymmteric bridge, one can observe that it only deforms the $\mathcal{F}$(C). Hence, if we construct any one of the seed correlator using the bridge, one can obtain all other correlators in the multiplet by using the SUSY. 

\textit{Example}: Using the $\mathcal{N}=2$ supersymmetric Grassmannian \cite{Bala:2026hdm} for the photon multiplet, we input the graviton correlator eq \eqref{eq:four-graviton-discussion} obtained using the bridge and obtain the photon four point with graviton exchange, 
\begin{align}
    A_{4,\text{photon}}^{--++}= &(12\bar3\bar4)^2 \frac{\Delta_S^2}{4 \Sigma^2}
\bigg[
\frac1S
\left(
\frac1{\Delta_U^2}
+
\frac1{\Delta_T^2}
\right)
\notag \\ &+
\frac1T
\left(
\frac1{\Delta_S^2}
+
\frac1{\Delta_U^2}
\right)
+
\frac1U
\left(
\frac1{\Delta_T^2}
+
\frac1{\Delta_S^2}
\right)
\bigg].
\end{align}
The above obtained correlator has the correct factorization and flat space limit.

Also note that if we use the scalar correlator, $\mathcal{A}_{4}$ \eqref{scalA4} obtained using the bridge, having only three poles S,T and $\Sigma$ as the seed to obtain the gluon correlator. One can observe that the non-compact geometry of the scalar correlator gets closed due to the helicity weight of gluon \cite{Bala:2026bdx} provided by SUSY, that promote the $\Delta=0$ to a pole, which closes the rectangle. \footnote{Similarly, using the scalar correlator as the seed, the four-gluino component has the schematic form
\[
A_{4}^{--++}
=
-{(12\bar 3\bar 4)\over 2\Sigma}
\left(
{1\over S}+{1\over T}
\right)
=
-{(12\bar 3\bar 4)(S+T)\over 2ST\Sigma}.
\]
Thus its poles are only at \(S=0\), \(T=0\), and \(\Sigma=0\); there is no pole at
\(\Delta=S+T-U=0\). In the \((x,y)\) variables this corresponds to the three
finite loci \(x=0\), \(x=1\), and \(y=-1\). This can be viewed as a
non-compact three-boundary geometry rather than the compact rectangle appearing
for the four-gluon component. One can write the corresponding non-compact canonical form as follows 
$$\Omega_\lambda
=
d\log {x\over 1-x}\wedge d\log(1+y)
=
{dx\wedge dy\over x(1-x)(1+y)}$$. }

\paragraph{\textbf{Discussion.}}The central result of this work is that the $O(n,n)$ bridge is an exact automorphism of the cosmological Grassmannian support. This reduces the four-gluon recursion to pure residue extraction — no spectral or radial integrals survive. The gluon answer can be interpreted as the canonical form of a compact rectangle. This  places factorization and energy singularities on identical geometric footing. The geometry can also be understood using the Pfaffians of the Grassmannian. We have to use multiple bridges in order to get all the channels of the graviton four point correlator. We also observed a double copy type relation of graviton four point away from the cut. The super-Grassmannian extension is equally clean, since the bridge preserves the fermionic support  $\hat\delta(C\cdot\Xi)$ alongside the bosonic one $\delta(C\cdot\Lambda)$. 
It is interesting to note that using SUSY, one can observe that it is the requirement of the gluon helicity weight that promotes the $\Delta$ in the scalar correlator to a pole, closing the quadrilateral into a compact positive geometry. This feature is absent in the scalar or the gluino component.

\paragraph{\textbf{Outlook.}}A promising next direction is to use supersymmetry more systematically. For the case of \(\mathcal N=4\) SUSY,  scalar correlators with graviton exchange could provide a simpler component from which the four-graviton correlator may be reconstructed by SUSY relations.  Another important test is the five-gluon correlator. This is not a straightforward extension of the four point case. At five points, the rank-one factorization rule that yields the four-gluon correlator on $\mathrm{OGr}(4,8)$ actually  over-constrains the off-diagonal block of $\mathrm{OGr}(5,10)$. It seems that  the five-gluon computation requires a rank-two gluing with a transverse modulus whose measure Jacobian must supply the precise internal little-group scaling needed  for  $\mathrm{GL}(1)$-invariance. Our ultimate aim is to produce the Parke-Taylor analog for AdS/dS correlators \cite{XYZ:2026new3}.
 It would also be very interesting to develop the BCFW for the higher-dimensional Grassmannian \cite{Bala:2026trw} correlator.
 \\
 
\textit{\textbf{Acknowledgment}}
AB acknowledges a UGC-JRF fellowship. We would like to thank Nipun Bhave and Atharva Abhyankar for useful discussions. We would especially like to acknowledge our debt to the people of India for their steady support of research in the basic sciences.










\bibliography{biblio}

@article{Britto:2004ap,
  author = {Britto, Ruth and Cachazo, Freddy and Feng, Bo},
  title = {New Recursion Relations for Tree Amplitudes of Gluons},
  journal = {Nucl. Phys. B},
  volume = {715},
  pages = {499--522},
  year = {2005},
  doi = {10.1016/j.nuclphysb.2005.02.030},
  eprint = {hep-th/0412308},
  archivePrefix = {arXiv},
  primaryClass = {hep-th}
}

@article{Britto:2005fq,
  author = {Britto, Ruth and Cachazo, Freddy and Feng, Bo and Witten, Edward},
  title = {Direct Proof of Tree-Level Recursion Relation in Yang-Mills Theory},
  journal = {Phys. Rev. Lett.},
  volume = {94},
  pages = {181602},
  year = {2005},
  doi = {10.1103/PhysRevLett.94.181602},
  eprint = {hep-th/0501052},
  archivePrefix = {arXiv},
  primaryClass = {hep-th}
}

@article{Raju:2010by,
    author = "Raju, Suvrat",
    title = "{BCFW for Witten Diagrams}",
    eprint = "1011.0780",
    archivePrefix = "arXiv",
    primaryClass = "hep-th",
    reportNumber = "HRI-ST-1009",
    doi = "10.1103/PhysRevLett.106.091601",
    journal = "Phys. Rev. Lett.",
    volume = "106",
    pages = "091601",
    year = "2011"
}

@article{raju2012new,
    author = "Raju, Suvrat",
    title = "{New Recursion Relations and a Flat Space Limit for AdS/CFT Correlators}",
    eprint = "1201.6449",
    archivePrefix = "arXiv",
    primaryClass = "hep-th",
    reportNumber = "HRI-ST-1201",
    doi = "10.1103/PhysRevD.85.126009",
    journal = "Phys. Rev. D",
    volume = "85",
    pages = "126009",
    year = "2012"
}

@article{PhysRevD.106.L121701,
  title = {New recursion relations for tree-level correlators in anti--de Sitter spacetime},
  author = {Armstrong, Connor and Gomez, Humberto and Lipinski Jusinskas, Renann and Lipstein, Arthur and Mei, Jiajie},
  journal = {Phys. Rev. D},
  volume = {106},
  issue = {12},
  pages = {L121701},
  numpages = {9},
  year = {2022},
  month = {Dec},
  publisher = {American Physical Society},
  doi = {10.1103/PhysRevD.106.L121701},
  url = {https://link.aps.org/doi/10.1103/PhysRevD.106.L121701}
}

@article{Arundine:2026fbr,
  title={The Cosmological Grassmannian},
  author={Arundine, Mattia and Baumann, Daniel and Lee, Mang Hei Gordon and Pimentel, Guilherme L and Rost, Facundo},
  journal={arXiv preprint arXiv:2602.07117},
  year={2026},
  eprint="Arxiv: 2602.07117"
}

@article{De:2026shn,
  title={The Vasiliev Grassmannian},
  author={De, Shounak and Lee, Hayden},
  journal={arXiv preprint arXiv:2603.24656},
  year={2026},
  eprint="Arxiv: 2603.24656"
}

@article{Bala:2026hdm,
  title={The $\mathcal{N}=1$ Super-Grassmannian for CFT$_3$ and a Foray on AdS and Cosmological Correlators},
  author={Bala, Aswini and Jain, Sachin and S., Dhruva K. and Rao, Adithya A.},
  journal={arXiv preprint arXiv:2604.07446},
  eprint="arXiv:2604.07446",
  year={2026}
}

@article{Bala:2026bdx,
  title={Super-Grassmannians for $\mathcal{N}=2$ to $4$ SCFT$_3$: From AdS$_4$ Correlators to $\mathcal{N}=4$ SYM scattering Amplitudes},
  author={Bala, Aswini and Jain, Sachin and S., Dhruva K. and Rao, Adithya A.},
  journal={arXiv preprint arXiv:2604.07503},
  eprint="arXiv:2604.07503",
  year={2026}
}

@article{Ansari:2025fvi,
    author = "Ansari, Arhum and Jain, Sachin and S, Dhruva K.",
    title = "{AdS$_{4}$ boundary Wightman functions in twistor space: factorization, conformal blocks and a double copy}",
    eprint = "2512.04172",
    archivePrefix = "arXiv",
    primaryClass = "hep-th",
    doi = "10.1007/JHEP06(2026)024",
    journal = "JHEP",
    volume = "06",
    pages = "024",
    year = "2026"
}

@article{Huang:2026tsh,
  title={Beyond Discontinuities: Cosmological WFCs from the Supersymmetric Orthogonal Grassmannian},
  author={Huang, Yu-tin and Kuo, Chia-Kai and Liu, Yohan and Mei, Jiajie},
  journal={arXiv preprint arXiv:2604.08512},
  eprint="arXiv:2604.08512",
  year={2026}
}

@article{Bala:2026trw,
  title={he Conformal Grassmannian: A Symplectic Bi-Grassmannian for $CFT_ 4$ Correlators},
  author={Bala, Aswini and Jain, Sachin and S, Dhruva K.},
  journal={arXiv preprint arXiv:2605.06811},
  eprint="arXiv:2605.06811",
  year={2026}
}

@article{Arundine:2026myr,
  title={Cosmological Collider in the Grassmannian},
  author={Arundine, Mattia and Pimentel, Guilherme L},
  journal={arXiv preprint arXiv:2605.21581},
  eprint="arXiv:2605.21581",
  year={2026}
}

@article{Arkani-Hamed:2017mur,
    author = "Arkani-Hamed, Nima and Bai, Yuntao and He, Song and Yan, Gongwang",
    title = "{Scattering Forms and the Positive Geometry of Kinematics, Color and the Worldsheet}",
    eprint = "1711.09102",
    archivePrefix = "arXiv",
    primaryClass = "hep-th",
    doi = "10.1007/JHEP05(2018)096",
    journal = "JHEP",
    volume = "05",
    pages = "096",
    year = "2018"
}

@article{Arkani-Hamed:2017fdk,
  title={Cosmological Polytopes and the Wavefunction of the Universe},
  author={Arkani-Hamed, Nima and Benincasa, Paolo and Postnikov, Alexander},
  journal={arXiv preprint arXiv:1709.02813},
  eprint="arXiv:1709.02813",
  year={2017}
}

@article{Arkani-Hamed:2024jbp,
    author = "Arkani-Hamed, Nima and Figueiredo, Carolina and Vaz{\~a}o, Francisco",
    title = "{Cosmohedra}",
    eprint = "2412.19881",
    archivePrefix = "arXiv",
    primaryClass = "hep-th",
    doi = "10.1007/JHEP11(2025)029",
    journal = "JHEP",
    volume = "11",
    pages = "029",
    year = "2025"
}

@book{Arkani-Hamed:2012zlh,
    author = "Arkani-Hamed, Nima and Bourjaily, Jacob L. and Cachazo, Freddy and Goncharov, Alexander B. and Postnikov, Alexander and Trnka, Jaroslav",
    title = "{Grassmannian Geometry of Scattering Amplitudes}",
    eprint = "1212.5605",
    archivePrefix = "arXiv",
    primaryClass = "hep-th",
    reportNumber = "PUPT-2435",
    doi = "10.1017/CBO9781316091548",
    isbn = "978-1-107-08658-6, 978-1-316-57296-2",
    publisher = "Cambridge University Press",
    month = "4",
    year = "2016"
}

@article{Huang:2013owa,
    author = "Huang, Yu-Tin and Wen, CongKao",
    title = "{ABJM amplitudes and the positive orthogonal grassmannian}",
    eprint = "1309.3252",
    archivePrefix = "arXiv",
    primaryClass = "hep-th",
    reportNumber = "QMUL-PH-13-09",
    doi = "10.1007/JHEP02(2014)104",
    journal = "JHEP",
    volume = "02",
    pages = "104",
    year = "2014"
}

@article{XYZ:2026new3,
    author = "",
    title ="Work in progress",
    journal = "",
    year = "20XX"
}

@article{XYZ:2026new4,
    author = "",
    title ="Work in progress",
    journal = "",
    year = "20XX"
}
\end{document}